\documentclass{llncs}

\begin{document}

\title{Counting Inversions Adaptively}
\author{Amr Elmasry} 
\institute{Department of Computer Engineering and Systems \\ Alexandria University, Egypt \\ elmasry@alexu.edu.eg}
\maketitle

\begin{abstract}
We give a simple and efficient algorithm for adaptively counting inversions in a sequence of $n$ integers. Our algorithm runs in $O(n + n \sqrt{\lg{(Inv/n)}})$ time in the word-RAM model of computation, where $Inv$ is the number of inversions. 
\end{abstract}

\section{Introduction}
Consider a sequence $X$ of $n$ elements drawn from a totally ordered set. 
The number of inversions in $X$ is defined as the number of element pairs in wrong order, i.e., 
$Inv(X) = |\{(i,j) ~ | ~ 1 \leq i < j \leq n \mbox{ and } X_{i} > X_{j}\}|$ \cite{k}. 
Obviously, $Inv(X) = 0$ if $X$ is sorted and can grow up to $n(n-1)/2$ if $X$ is inversely sorted. 
A folk algorithm for counting inversions is to embed a subroutine that keeps track of the number of inversions within the Merge-sort algorithm. It is thus well known that one can optimally count the inversions for a sequence of $n$ elements in $O(n \lg n)$ time in the comparison-based model of computation. 

Sorting algorithms can benefit from the presortedness in the input.
In the comparison-based model, an adaptive sorting algorithm is optimal with respect to the number of inversions when it runs in $O(n + n\log(Inv/n))$ time \cite{gmpr}.
There are plenty of adaptive sorting algorithms that achieve this bound \cite{bt,ef,e1,e2,e3,ew,lp,m79}.
It is almost always possible to augment the underlying data structures of those adaptive sorting algorithms  
with counters to keep track of the number of inversions as a byproduct.
It then follows that inversions counting can be adaptively performed in $O(n + n\log(Inv/n))$ time.

When it comes to the word-RAM model, there exist integer sorting and inversions-counting algorithms that are asymptotically faster.
The best known deterministic integer sorting algorithm runs in $O(n \lg \lg n)$ time \cite{h}, 
and the best known randomized integer sorting algorithm runs in $O(n \sqrt{\lg \lg n})$ expected time \cite{ht}.
The inversions counting problem is apparently more difficult than sorting in this model;
the running time of the best known inversions-counting algorithm is $O(n \sqrt{\lg n})$ \cite{cp}.
For sorting, adaptive counterparts run in $O(n + n \lg \lg (Inv/n))$ worst-case time and $O(n + n \sqrt{\lg \lg (Inv/n)})$ expected time \cite{ppt}.

In this paper we give an adaptive algorithm that counts inversions in $O(n + n \sqrt{\lg(Inv/n)})$ time in the word-RAM model of computation.
More generally, for any function $f(.)$, deploying an $O(n \cdot f(n))$-time subroutine for inversions counting, our algorithm runs in $O(n + n \cdot f(Inv/n))$ time.

\section{The algorithm}
Our algorithm maintains a data structure comprising a list of lists. 
The elements of each list are smaller than or equal to the elements of the subsequent lists; 
the elements within a list are not necessarily sorted though. 
We maintain at the header of each list the count of elements in the list and the value of its maximum element.
For a global non-decreasing parameter $q$ initialized to one, the number of elements in each list is between $q$ and $2q$,
except possibly for the last list that may have fewer elements.
We also keep track of an inversions count that is kept in a global variable initialized to zero.

Handling the input sequence in reverse order, in each iteration a new element is inserted into the list of lists.
The insertion is done by scanning the headers of the lists sequentially, and comparing the current element with the maximum of each list.
The current element is inserted at the beginning of the first list whose maximum is not smaller than the element. 
Otherwise, if the current element is larger than all the already inserted elements, it is inserted at the beginning of the last list, whose maximum is accordingly updated. While passing by the header of a list, the count of its elements is added to the inversions count.

If, following the insertion, the number of elements in one list becomes $2q+1$, the list is split in two almost equal lists. 
This is done by applying a median-finding algorithm, keeping the $q+1$ elements smaller than or equal to the median in the same old order in the current list, forming a new list from the other $q$ elements in the same old order, and inserting the new list next to the other one.
At this point, we need to account for the found inversions as a result of this split. 
One way to do that efficiently is to keep track of the priori position $p$ of each element in the list before the partitioning and its position $\pi(p)$ in the resulting list partitioned around the median just before the split. We then add the sum $\sum_{\pi(p)>p} (\pi(p) - p)$ to the accumulated inversions count.

The key point for making the algorithm adaptive is controlling the parameter $q$.   
The algorithm works in phases, where in the $i$th phase the value of $q$ is set to $q_i=2^{i-1}$.
Let $t_i$ be the number of elements inserted in the $i$th phase.
Once the number of comparisons performed in the $i$th phase, for comparing the inserted elements in this phase with the headers of the lists, exceeds $n/ \sqrt{q_i} + t_i$, we conclude that our estimate for the number of inversions is smaller than what it should be. In accordance, we interrupt the current insertion, double the value of $q$, and begin a new phase. For that, we reorganize the data structure by combining every other list with the following list, appending the latter at the tail of the former, and updating the header information for each combined list.  
If the number of lists was initially odd, we leave the last list alone.

By inserting all the elements in the data structure, we have accounted for the inversions across the current lists but not for the inversions within the lists. Recall that the elements within each list are not yet sorted. Our algorithm concludes by calling the algorithm of Chan and 
P\v{a}tra\c{s}cu \cite{cp} for counting inversions within each list, and adding these counts to the accumulated inversions count.

\section{Analysis}

First, we argue that the algorithm terminates by guaranteeing that all the elements are inserted in the data structure.
The number of lists in the $i$th phase is at most $\lceil n/2^{i-1} \rceil$. Hence, unless all the elements are inserted beforehand, 
there will only be one list after at most $\lceil \lg n \rceil + 1$ phases. In such case, only one comparison is need to insert 
each element in the list. The condition for terminating a phase will then not be fulfilled until all the elements are inserted.

To demonstrate the correctness of the algorithm we need to show that it counts the inversions correctly.
We adopt the strategy of counting inversions while fixing them. In other words, whenever the inversions accompanying an element are fixed, putting 
the element in order, the number of inversions that this element is involved in are added to the accumulated inversions count.

While inserting an element, as the elements are inserted in reverse order, all the elements in the lists, whose headers are passed by, count towards the number of inversions. Indeed, we add the number of elements of these lists to the total number of inversions. 
Every element is then inserted at the beginning of the list where it belongs. The next action that may affect the order of the elements is when a list is partitioned around its median. We explicitly account for the inversions revealed in this partitioning by subtracting the old position of each element from its new position, if the latter is larger than the former. 
To illustrate the correctness of this calculation, denote the elements that land in the second half as {\it big} and those in the first as {\it small}. Note that the order among big elements and the order among small elements is not altered by the partitioning.  
The difference between the new position of a big element and its old position is precisely the number of small elements that appeared after this big element in the original list, and are naturally moved before it after the partitioning. These are precisely the number of inversions that this big element is involved in. The sum of these counts is indeed the total number of the revealed inversions. On the other hand, for all the small elements $\pi(p)$ is less than or equal to $p$, and hence are not accounted for in the formula. 
At the end of this part of the algorithm, all the inversions across the lists must have been accounted for. 

To analyze the running time, we separately consider four basic actions:

\begin{enumerate}
\item the work done during the insertions to compare the inserted elements with the headers of the lists,
\item the work done for splitting the lists using the median-finding subroutine,
\item the work done at the end of each phase for combining pairs of lists, and
\item the work done by the Chan-P\v{a}tra\c{s}cu algorithm.
\end{enumerate}

For the first type, the total number of comparisons performed in the $i$th phase is at most $n/\sqrt{q_i} + t_i +1$. 
Since in the $i$th phase $q_i=2^{i-1}$, the first-type work is proportional to $\sum_{i=1}^{\lceil \lg n \rceil +1} (n/\sqrt{2^{i-1}} + t_i +1) = O(n)$.
In words, the work to compare the inserted elements with the headers of the lists is linear in $n$.

For the second type, we use the accounting method to amortize the work done. 
We charge a constant number of credits to every inserted element. 
We maintain on each list a number of credits that is equal to its number of elements minus $q$.
When a new element is inserted into a list, the credits on the element are moved to the list.
When the size of the list is $2q$, the credits on the list are $\Theta(q)$ and are enough to pay for the
work done in the median-finding and splitting process. Consider the case when a new phase is in action. 
The value of $q$ then doubles becoming $q'=2q$. Consider any two lists of lengths $\ell_1$ and $\ell_2$ that are combined in this phase. 
There must have been $\ell_1 - q$ and $\ell_2 - q$ credits on the two lists respectively. 
The combined list now has length $\ell_1 + \ell_2$ and the number of credits on it is $\ell_1 + \ell_2 - 2q = \ell_1 + \ell_2 - q'$,
fulfilling the accounting requirement. In conclusion, the amount of work done during splitting the lists and calling the median-finding subroutine 
is also linear in $n$.

For the third type, we note that the work done to combine two lists is a constant. 
Also, there are at most $\lceil \lg n \rceil +1$ phases, and the number of lists in the $i$th phase is at most $\lceil n/2^{i-1} \rceil$.
Hence, the work done in combining the lists is at most $\sum_{i=1}^{\lceil \lg n \rceil +1} \lceil n/2^{i-1} \rceil = O(n)$. 

For the fourth type, we need to bound the final value, $\hat{q}$, of $q$ in terms of the total number of inversions, $Inv$.
Consider the phase just before the last phase, where $q$ is equal to $\hat{q}/2$.
Since in this phase the number of comparisons performed for comparing the inserted elements with the headers of the lists 
exceeds $n/\sqrt{\hat{q}/2} + t_i$, and since the number of elements per list is at least $\hat{q}/2$,
then $Inv \geq n \sqrt{\hat{q}/2}$. Hence, $\hat{q} \leq 2 \cdot (Inv/n)^2$. The size of each list in the last phase is at most $2\hat{q}$, which is then $O((Inv/n)^2)$. Applying Chan and P\v{a}tra\c{s}cu's algorithm on one list requires $O(\hat{q} \sqrt{\lg{\hat{q}}})$ time.
It follows that the total work done to count the inversions within all the lists is $O(n \sqrt{\lg{\hat{q}}}) = O(n \sqrt{\lg(Inv/n)})$.
           
Our main theorem follows

\begin{theorem}
The number of inversions, $Inv$, in a sequence of $n$ integers, can be counted in $O(n + n \sqrt{\lg(Inv/n)})$ time in the word-RAM model of computation.
\end{theorem}

\end{document}